\documentclass{mem}
\usepackage{natbib}
\usepackage{balance}
\usepackage{graphicx}
\usepackage{flushend}
\usepackage{xcolor}
\usepackage[version=4]{mhchem}
\usepackage[breaklinks,pdftex]{hyperref}
\usepackage{txfonts}

\idline{00}{001}
\begin{document}

\title{Ab initio Calculations for Astrochemistry
}


\author{
F. \,Tonolo
\and S. \,Alessandrini
}

\institute{Dipartimento di Chimica ``Giacomo Ciamician", Alma Mater Studiorum - Università di Bologna, via Francesco Selmi 2, Bologna, 40126, Italy \\
\email{francesca.tonolo@sns.it, silvia.alessandrini7@unibo.it}
}

\authorrunning{Tonolo}
\titlerunning{Ab initio Calculations for Astrochemistry}

\date{Received: XX-XX-XXXX; Accepted: XX-XX-XXXX }

\abstract{
Computational chemistry plays a relevant role in many astrochemical research fields, either by complementing experimental measurements or by deriving parameters difficult to be reproduced by laboratories.
While the role of computational spectroscopy in assisting new observations in space is described, the core of the chapter is the investigation of the collisional radiative transfer and the bimolecular reactive processes occurring in the gas-phase conditions of the interstellar medium, using as a guide the contributions presented by the authors at the ``Second Italian National Congress on Proto(-planetary) Astrochemistry'', held in Trieste in September 2023. In particular, the need for accurate datasets of collisional coefficients to model molecular abundances will be discussed. Then, the role of quantum chemistry in the investigation of interstellar-relevant potential energy surfaces will be described, focusing on accurate thermodynamic quantities for the estimate of rate coefficients.

}
\maketitle{}

\section{Introduction}

\begin{figure*}
    \centering
    \includegraphics[scale=0.25]{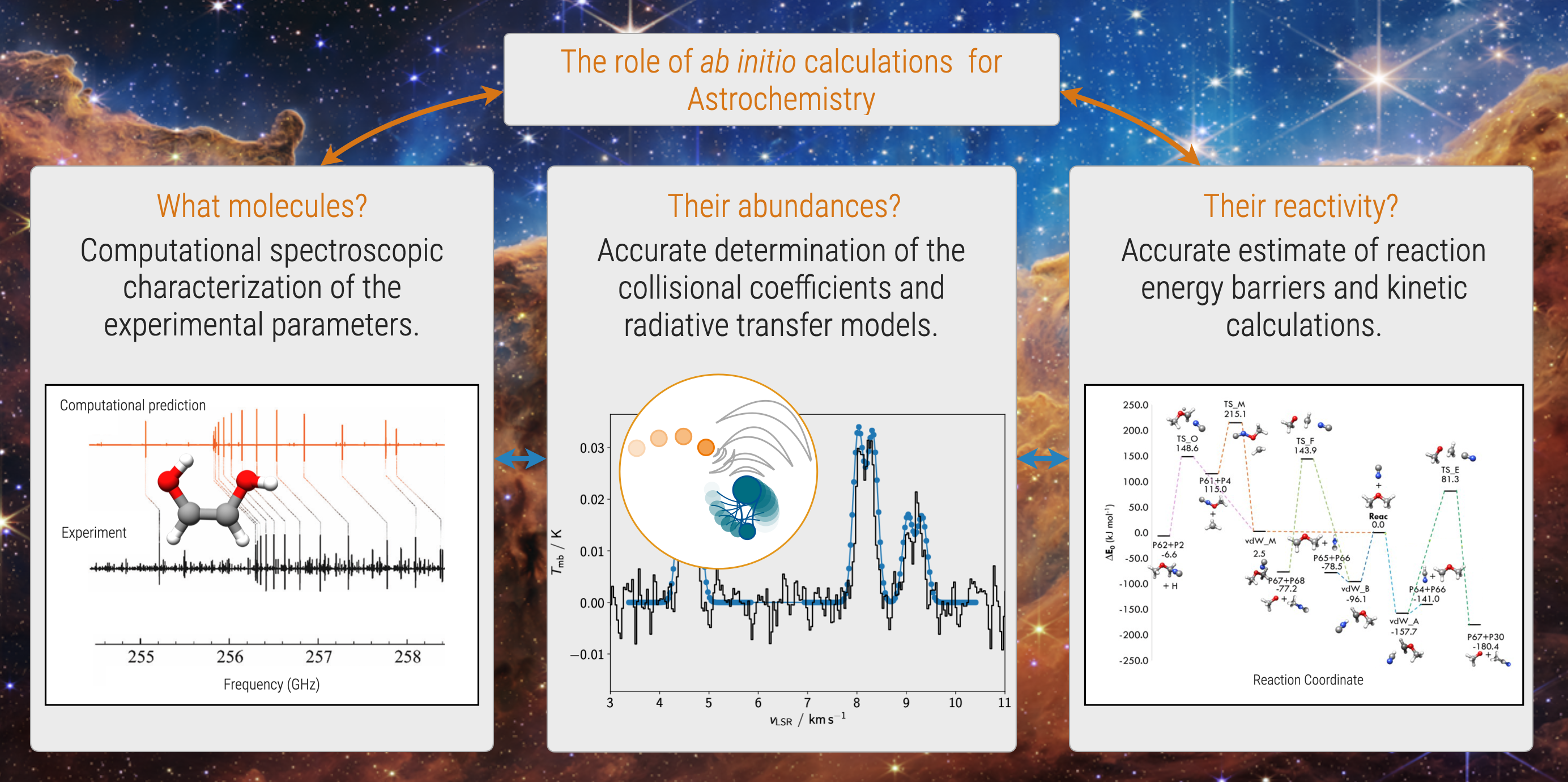}
    \caption{The role of ab initio calculations in astrochemistry: from the interplay between theory and experiment in rotational spectroscopy to the derivation of molecular abundances and gas-phase reaction profiles for kinetic calculations.}
    \label{fig:Fra}
\end{figure*}
Quantum-chemical approaches are a powerful tool for astrochemists, with 
ab initio methodologies widely
used to obtain experimentally inaccessible information and insights into processes occurring in the gas phase of the interstellar medium (ISM).
Here, the temperature can reach values of a few tens of K and densities are usually between $10^2$ and $10^6$ molecules$\cdot$cm$^{-3}$, thus posing constraints on both reactive and dynamic processes \citep{tielens2013molecular,yamamoto2017introduction}. However, the nature of quantum-chemical calculations in describing isolated molecules well matches with the extremely low-density environment of the gas phase, where molecules are mostly observed.

Observations of molecules in the ISM are based on radioastronomical techniques and these have flourished in the recent years thanks to facilities like the ALMA \citep{wootten2009atacama}, Yebes \citep{tercero2021yebes}, IRAM \citep{baars1987iram} and GBT \citep{white2022green} telescopes and the associated large surveys\citep{mcguire2020early,Pepe2022Conf}. 
While the present chapter does not consider ices and reactions occurring on their surfaces, as the main focus is on gas-phase processes, it should be mentioned that the recent launch of the JWST started also to provide observations of icy dust mantles by means of the vibrational signals in the infrared region. However, the species identified are still very simple, like \ce{CH3+} \citep{berne2023formation,mcclure2023ice}.

The radioastronomical detection of a molecule relies on the observation of its rotational transitions, which should be known with great accuracy by means of a thorough experimental work. Only a few species, like \ce{C5N-} \citep{cernicharo2008detection}, have been observed based only on their accurate theoretical predictions. This is due to two reasons. First, the simulation of rotational spectroscopic parameters is strongly affected by the level of theory employed \citep{puzzarini2023connections,puzzarini2010quantum,alessandrini2018accuracy}. For example, rotational constants, the leading term of rotational spectroscopy, obtained with an error of about 0.1\% with respect to the experimental data, translates into computed geometries with an accuracy of 0.0005–0.001 \AA\ \citep{puzzarini2023connections}. Secondly, the rotational transitions simulated with the rotational constants are then affected by several interactions, among which the vibration-rotation one, always present. 
These interactions shift the computational predictions by an unknown quantity, prohibiting the direct comparison with radioastronomical data. Computational methods able to reproduce the experimental results within the accuracy mentioned above are computationally affordable only for very small molecules (up to 5/6 atoms), thus they cannot be employed for complex-organic molecules, $i.e.$, the main target of modern astrochemistry. In this regard, a large effort has been made by theoretical chemists to propose new tools which lower the computational cost and still retain a good accuracy. For example, composite schemes for energies and proprieties \citep{puzzarini2010quantum,heckert2005molecular,tajti2004heat}, or explicitly correlated F12 methods \citealt{gruneis2017perspective}). Efforts have been made also in the implementation of new techniques, like the Cholesky decomposition \citep{aquilante2011cholesky,nottoli2021black}. In addition, the possibility of adopting ``experimentally accurate'' data to correct computational quantities has been used to improve the spectroscopic predictions of larger species, like PAHs or long carbon chains \citep{melli2021extending,ye2022exploiting,puzzarini2023hunting}. Nevertheless computational spectroscopic quantities cannot be employed to analyze radioastronomical data most often, but they have a prominent role in rotational spectroscopy. Indeed, they are used to guide experimental measurements of unstable species \citep{melosso2022gas,puzzarini2023hunting}, not known in the literature, and to constrain parameters that the experiment is not able to determine \citep{cazzoli2014rare}.

The role of accurate computational methods does not end with rotational spectroscopy and assumes a prominent role in the derivation of accurate collisional coefficients and kinetic estimates of gas-phase neutral-neutral reactions. As illustrated in Figure \ref{fig:Fra}, the previous quantities are fundamental to derive molecular abundances and build reaction networks, respectively. 
The outcomes of these investigations are not stand-alone results, but their interplay is essential to support the physical and chemical modeling of the ISM. For an extended description of the most relevant processes occurring in the ISM, the reader is referred to key reviews and books such as \cite{dalgarno1976molecule,watson1976interstellar,tennyson2003molecules,wakelam2010reaction,bates2012atomic} and \cite{yamamoto2017introduction}. In addition, to 
ease the interplay of these investigations for astrochemical modeling purposes, the available information is now centralized in several public databases. 
To cite a few: the BASECOL \citep{dubernet2024basecol2023}, LAMDA \citep{schoier2005atomic}, and EMAA (\url{https://emaa.osug.fr//}) databases furnish the collision dynamics outcomes. For kinetics rate constants, the KIDA \citep{wakelam2012kinetic} and UMIST \citep{millar2024umist} databases are the most widely known, while for spectroscopy different sets have been developed depending on the frequency range. In particular, the CDMS database collects rotational lines \citep{endres2016cologne}, the LIDA one IR bands of molecules in ice mixtures \citep{rocha2022lida}, EDIBLES \citep{cox2017eso} diffuse interstellar bands, and ExoMol \citep{tennyson2016exomol} is specifically contaning line list useful for the modeling of exoplanets and hot atmospheres.

For more details on computational spectroscopic parameters and their accuracy the reader is referred to \cite{puzzarini2010quantum} while in the following collision dynamics and reactive potential energy surfaces (PES) will be addressed. Section \ref{colldyn} will focus on the accurate determination of collisional coefficients that are used to model the abundances of molecules observed in the ISM, where local thermodynamic equilibrium (LTE) conditions are rarely fulfilled. The last part of the chapter (section \ref{reac}) will explore the role of ab initio methods for the derivation of accurate kinetic rate constants of gas-phase for neutral-neutral reactions. 

In the following sections, the examples provided are the ones presented during the ``Second Italian National Congress on Proto(-planetary) Astrochemistry'', held in Trieste in September 2023. 

\section{The role of collision dynamics calculations to model molecular abundances in space}
\label{colldyn}
The derivation of molecular abundances in space needs to account for the sparse physical conditions that can be found in astrophysical environments. Indeed, the radiative transfer equations used to model astrophysical observations and derive molecular column densities exhibit large sensitivity to the processes that affect the population distributions among molecular levels. Such processes strongly depend on the physical conditions of the targeted environment.  
For example, in the ISM the density is so low ($\sim10^2-10^6$\,molecule$\cdot$cm$^{-3}$) that 
molecular energy level populations are not in LTE. Under such conditions, the derivation of molecular abundances from spectral lines requires the knowledge of the collisional rate coefficients of the molecule under consideration for the most abundant perturbing species, $i.e.$, \ce{H2}, \ce{H} and \ce{He} \citep{roueff2013molecular,lique2019gas}. 
The variation in accuracy of the collisional rate coefficients can cause differences up to a factor of 10 in the line intensities, which reflects to a significant change in the prediction of the molecular abundances ($e.g.$, \citealt{sarrasin2010rotational,lanza2014new}). Therefore, the progress in radiative transfer calculations needs to go hand in hand with the improvement in the accuracy of the collisional predictions. This is also reflected in the major efforts that are currently devoted to reproduce these coefficients from an experimental point of view \citep{yang2010communication,yang2011state, chefdeville2012appearance,brouard2014taming,bergeat2015quantum,bergeat2020probing}. 
Nowadays, given the paucity of experimental setups able to probe collision dynamics, the knowledge of the collisional rate coefficients strongly relies on theoretical calculations \citep{roueff2013molecular,lique2019gas}. 
Thus, an accurate yet affordable computational procedure for the characterization of the collisional properties of astrochemical molecules, and subsequent modeling of their spectroscopic transitions in terms of abundance, needs to be validated. 
In the following, an illustrative protocol that is particularly suited to inspect the collisional behavior of small ionic systems in both a reliable and undemanding manner is presented.
The latter consists of four main steps.

The first one is the investigation of the collisional PES by means of ab initio calculations. This step requires particular attention because uncertainties in the PES have a significant impact on the accuracy of the collisional parameters (see \citealt{faure2021collisional} for an illustrative example). Hence, high computational accuracies need to be achieved. 
This usually exploits the excellent performances of explicitly correlated coupled cluster methods \citep{adler2007simple,knizia2009simplified} to describe interaction energies \citep{ajili2013accuracy,tonolo2021improved}. When ionic systems are involved, the use of fully augmented basis sets is particularly promising in describing the energetics of the long-range regions of the potential, where dispersive interactions are more relevant \citep{kendall1992electron}. 

The second step consists of expressing the potential as an expansion over angular functions. This often requires to resort to some approximations to reduce the computational cost of scattering calculations. This applies, in particular, to systems involving a collisional partner with a rotational structure, such as \ce{H2}. In these cases, the effects due to the coupling between the different rotational states of the collider (indexed using the $J$ quantum number) on the inelastic cross sections need to be preliminarily assessed. 
This brings important hints on the feasibility of neglecting the $J>0$ rotational states of the collider (the so-called ``spherical approximation''). 
The reduced computational cost of this approximation, if properly validated, permits to extend this procedure to larger molecular systems and to a wide range of astrochemical conditions. 
The spherical approximation has been proved to be particularly suited to small ionic systems interacting with \ce{H2}
\citep{spielfiedel2015new, balancca2020inelastic, cabrera2020relaxation}. 
For instance, for the \ce{HC^{17}O+} and \ce{PO+} targets \citep{tonolo2022hyperfine,tonolo2023collisional} the impact of the $J>0$ rotational levels of \ce{H2} resulted to be considerably weak (the inclusion of the $J=2$ state of \ce{H2} led to average deviations of $\sim7$\% in the values of the cross sections, while the cross sections of \emph{para}-\ce{H2} and \emph{ortho}-\ce{H2} agreed within $\sim12$\% on average).

The third step is the solution of the close-coupling scattering equations in a range of energies that allows to derive the corresponding collisional parameters in the conditions of interest. 
For example, the pressure broadening and pressure shift parameters can be computed and can be subsequently used to infer the quality of the potential (see, as an example, \citealt{tonolo2021improved}).
From the scattering calculations, the inelastic state-to-state rate coefficients are also obtained.

The last step consists in using the collisional rate coefficients to model the rotational transitions observed in the interstellar environments by means of radiative transfer calculations.  
For example, the computed collisional dataset for the \ce{PO+}/\ce{H2} system helped to 
test the reliability of the LTE approximation and to refine the column density value of \ce{PO+} obtained from the observations of the G+0.693–0.027 molecular cloud \citep{rivilla2022ionize}.
Additionally, radiative transfer calculations indicated maser behavior for the first rotational transitions of \ce{PO+} at various kinetic temperatures and densities typically found in interstellar sources. This emphasizes the importance of accurate collisional coefficients for the precise modeling of molecular abundances in the ISM.

\section{Simple organics from gas-phase reactions: a computational view}
\label{reac}

The observation of new small organic molecules claims for innovative formation routes either on grains or in the gas phase of the ISM \citep{mcguire20222021,puzzarini2022gas}. The latter class of reactions is challenging due to the prohibitive conditions of the ISM and the three main points that guide neutral-neutral gas-phase reactions are: (i) only bi-molecular reactions can occur, and due to the lack of a third-body stabilization only bi-molecular products can be formed\citep{yamamoto2017introduction}; (ii) no external energetic input is provided for the process, so the products must be exothermic and have to involve submerged barrier with respect to the reactants; (iii) to make a collision reactive, the process has to involve a radical species, like \ce{CN}, \ce{NH} or \ce{CH}. Still, also other gas-phase processes occur in the ISM, like photodissociation or ion-molecule reactions \citep{yamamoto2017introduction,tielens2013molecular}.

Based on these three guidelines, it is possible to select reactions based on thermodynamic estimates and move to the kinetic study only for a small selection of processes. Indeed, the critical step in the gas-phase chemistry of the ISM is the estimation of rate coefficients that should be at least in the order of $10^{-11}/10^{-12} $ cm$^{3}$ molecule$^{-1}$ s$^{-1}$ \citep{yamamoto2017introduction,tielens2013molecular}. The latter can be obtained from experimental set-ups or using first principle computational methodologies. In both cases, limitations are involved: experimental measurements can suffer from the impossibility of reproducing interstellar-medium-like conditions. At the same time, theoretical kinetic estimates are strongly affected by the energetic quantities employed for their calculation and the approximation introduced in the procedure. A discussion on the accurate derivation of rate constants is out of the scope of the present chapter, but the need for accurate reaction barrier heights has to be mentioned and it is where ab initio computations play the main role. Indeed, meaningful kinetic rate constants can be obtained only if post-Hartree Fock methods, like coupled-cluster (CC) methodologies \citep{shavitt2009many} or multi-reference (MR) formulations, like MR configuration interaction (MRCI) \citep{buenker1974individualized} are used. In the case of CC techniques, one has to refer to the CCSD(T) method that includes singles and doubles excitations and a perturbative treatment of triples excitations \citep{stanton1997ccsd}\citep{zheng2009dbh24,klippenstein2017theoretical}. For example, it has been pointed out that the use of hybrid functional led to the wrong theoretical outcome in the case of the reaction between \ce{CH3CN} + \ce{CN} \citep{sleiman2018low,puzzarini2020twist} and in this particular case the CCSD(T) method in conjunction with standard basis set was not able to reproduce more accurate methodologies \citep{lupi2020state}. 

However, the accuracy can be pushed to its nowadays limit by employing explicitly correlated F12 methodologies \citep{adler2007simple,knizia2009simplified} in combination also with composite schemes, where several terms are computed at the best compromise between accuracy and computational cost and then combined to reach a better estimate of a property, like the energy \citep{tajti2004heat,barone2021development,ventura2021svecv,alessandrini2019extension,lupi2021junchs,barone2023reliable}. If accurate methods are employed, the reference geometry for the calculations can be obtained from density functional methods \citep{zheng2009dbh24}, either using hybrid or double-hybrid functionals \citep{becke1988density,kohn1965self,santra2019minimally}. This is a good compromise between accuracy and cost, considering DFT geometries provide structures that are as accurate as those obtainable from more expensive methods, like the CCSD(T) one, but can be employed on extremely vast PES, like those mapped for the \ce{CH} radical \citep{nikolayev2021theoretical,he2022chemical}. 

The use of computational means has also highlighted the possibility of studying several reaction mechanisms between the same stable species and different radicals. This is useful to understand if common reaction paths occur in the ISM, an hypothesis was first done for methanimine that is considered the precursor of complex imines in the ISM \ce{CH2NH} \citep{puzzarini2021chemical,barone2022toward,puzzarini2022gas}. The reaction between methanimine and \ce{CN} was analyzed by \cite{vazart2015cyanomethanimine} and leads to cyanomethanimine, (\ce{HNC=CHCN} as well as \ce{CH2NCN} species), all observed in the ISM \citep{zaleski2013detection,san2024first}. The same reaction with the \ce{CCH} radical forms propargylimine (HNC=CHCCH), another molecule detected in recent years \citep{bizzocchi2020propargylimine}, while the \ce{OH} radical seems to lead to the formation of formamide \citep{vazart2016state,rubin1971microwave}. Do the \ce{CH2NH + CN}, \ce{CH2NH + CCH} and, \ce{CH2NH + OH} reactions have something in common? The PES involves a radical in the doublet electronic state and points to identical exothermic reaction paths with slightly different energy barriers (even if still submerged). To confirm the presence of a general reaction mechanism, the processes occurring between \ce{CH2NH} and a large set of radicals having an unpaired electron (\ce{HCO, HCS, SH, NO, NS, SiN, C3N, and CP}) were analyzed \citep{alessandrini2021search,ye2024general}. Both H-abstraction and radical addition were considered and the products were hypothesized using the general reaction mechanism. Thus, the first step was the derivation of accurate energetic quantities to asses whether exothermic products are formed. This step was carried out using double-hybrid functional for geometries and vibrational frequencies (harmonic) then, the junChS \citep{alessandrini2019extension} composite scheme for the energies. Only the \ce{C3N} and the \ce{CP} radicals can lead to an exothermic path after the addition of the radical, while open channels for H-abstraction seem to occur for these species, but also for \ce{SH} and \ce{HCS}. For all the exothermic paths, the full potential energy surface has been explored with the double-hybrid functional and the energy was refined with the junChS composite scheme. A common reaction mechanism was observed for radical addition and H-abstraction, confirming the validity of the general mechanism.

Still, kinetics estimates are those able to point out if a reaction in the ISM is feasible. In the case of the methanimine-plus-radical reactions, the kinetic estimate where obtained using variational transition state theory with the Master Equation System Solver program (MESS, \citealt{georgievskii2013reformulation}) and using phase space theory for the derivation of the the kinetic term for the barrierless approach. According to our results, the processes leading to the addition of the radical on the C atom of \ce{CH2NH} are fast even in the ISM conditions, thus further pointing out \ce{HNCHCP} and \ce{HNCHC3N} species might be present in the ISM and only laboratory characterization might be missing for their observation. Even if the work here discussed employs statistical approached to obtain reaction rates, it should be mentioned that other routes are available to derive the rate coefficients. Among them, quantum dynamics is the reference strategy if one aims at experimental accuracy based on theoretical means. However, this method can only be used for systems with a reduced dimension, like the destruction of \ce{CH+}\citep{bovino2015ch+} or the reaction \ce{H + H2} \citep{ghosh2021charge}.

\section{Conclusions}
In this chapter, an overview of the role of ab initio strategies as a support of gas-phase astrochemical investigations is presented, from the prediction of the spectroscopic parameters to assist the experimental characterization of interstellar molecules, to reactivity and collision dynamics calculations. Specifically, the authors brought into focus the latter two topics and the results presented during the ``Second Italian National Congress on Proto(-planetary) Astrochemistry''. First, a description of the various steps required to compute the collisional coefficients and derive non-LTE abundances of interstellar molecules has been outlined. 
Secondly, we have discussed how the investigation of general mechanisms might be useful to describe interstellar chemistry and suggest new interstellar molecules to observe. In doing so, the need for accurate ab initio methodologies to derive reliable kinetics rate coefficients was mentioned. 
 
\begin{acknowledgements}
Both authors acknowledge the PRIN grant No. 202082CE3T ``ARES-A Road from Earth to the Stars'' for financial support and the ROT\&Comp Lab for the inspiring discussions over the past years. 
\end{acknowledgements}

\bibliographystyle{aa}
\bibliography{bibliography}

\end{document}